\documentclass[aps,prd,eqsecnum,showpacs]{revtex4}
\usepackage{graphicx}
\usepackage{amsfonts}
\usepackage{amssymb}
\begin{document}

\thispagestyle{empty}
\title{Semiclassical Instability of the Cauchy Horizon in Self-Similar Collapse}

\author{
Umpei Miyamoto$^{1}$
\footnote{Electronic address:umpei@gravity.phys.waseda.ac.jp}}
\author{
Tomohiro Harada$^{2}$
\footnote{Electronic address:T.Harada@qmul.ac.uk}}

\address{
$^{1}$ Department of Physics, Waseda University,
Okubo 3-4-1, Shinjuku, Tokyo 169-8555, Japan
}
\address{
$^{2}$ Astronomy Unit, School of Mathematical Sciences,
Queen Mary, Mile End Road, London E1 4NS, UK
}

\date{\today}

\begin{abstract}                
Generic spherically symmetric self-similar collapse results in
strong naked-singularity formation. 
In this paper we are concerned with particle
 creation during a naked-singularity formation in spherically symmetric
self-similar collapse without specifying the collapsing matter. 
In the generic case, the power of particle emission is found to be proportional to 
the inverse square of the remaining time to the Cauchy horizon (CH).
The constant of proportion can be arbitrarily large in the limit
 to marginally naked singularity.
Therefore, the unbounded power is especially striking in the case that
an event horizon is very close to the CH because
the emitted energy can be arbitrarily large in spite of
a cutoff expected from quantum gravity.
Above results suggest the instability of the CH in spherically symmetric 
self-similar spacetime from quantum field theory and seem to support the existence
of a semiclassical cosmic censor.
The divergence of redshifts and blueshifts of emitted particles
is found to cause the divergence of power to positive or negative infinity, depending
on the coupling manner of scalar fields to gravity.
On the other hand, it is found that there is a special class of self-similar spacetimes
in which the semiclassical instability of the CH 
is not efficient.
The analyses in this paper are based on the geometric optics approximation,
which is justified in two dimensions but needs 
justification in four dimensions.
\end{abstract}

\maketitle

\section{introduction}

The cosmic censorship hypothesis asserts that gravitational collapse of physically reasonable matter with generic initial condition does not result in the formation of naked singularity (NS)~\cite{PenrosePenrose2}. 
More precisely, there are two versions of this hypothesis. The weak hypothesis states that all singularities in gravitational collapse are hidden within black holes. This implies the future predictability of the spacetime outside the event horizon. The strong hypothesis asserts that no singularity visible to any observer can exist. It states that all physically reasonable spacetimes are globally hyperbolic.
Although the hypothesis has been an object of study over the last few decades,
there is little agreement as to the final end of a general relativistic gravitational collapse.
There is not enough evidence to prove the hypothesis.
On the contrary, some solutions of the Einstein equation with regular initial conditions evolving into spacetimes which contain naked singularities have been found.

Up to now, it has been reported that such semiclassical effects as particle creation by
naked-singularity formation, which are analogous to the Hawking
radiation~\cite{Hawking}, might prevent the formation of NS.
The first researchers to give much attention to this possibility were Ford and Parker.
They considered a globally naked-singular spacetime, in which the weak hypothesis is broken,
and calculated quantum emission from a
shell-crossing singularity to obtain a finite amount of 
flux~\cite{Ford}. On the other hand, Hiscock, Williams, and Eardley obtained 
diverging flux from a shell-focusing NS which results from a self-similar
implosion of null dust~\cite{Hiscock}. Subsequently, 
these semiclassical phenomena have been investigated in the collapse of
the self-similar dust~\cite{BarveBarve2,VazWitten}, non-self-similar
analytic dust \cite{HIN1HIN2}, and self-similar null-dust fluids ~\cite{Singh}.
Typically, the power of quantum emission from shell-focusing NSs 
diverges as the Cauchy horizon (CH) is approached.
For a recent review of the quantum/classical emission during naked-singularity formation, see~\cite{HIN3}.
In addition, such an explosive radiation by naked-singularity formation can be a candidate
for a source of the ultra high energy cosmic rays or a central engine of
$\gamma$-ray burst~\cite{GRB}.

As the examples given above show, it is known that generic spherically symmetric self-similar 
collapse results in strong naked-singularity formation~\cite{WaughLake,LakeZannias}.
Among such self-similar models,
the general relativistic Larson-Penston (GRLP) solution
would be one of the most serious counterexamples against 
the cosmic censorship hypothesis in the sense that
the existence of pressure is taken into account~\cite{OriPiranOriPiran2,OriPiran3}.
Moreover, the convergence of more general spherically symmetric 
collapse to the GRLP solution have been reported both 
numerically and analytically~\cite{HaradaMaeda} 
as a realization of the self-similarity hypothesis
proposed by Carr~\cite{Carr,CarrColey_1}. 
The discovery of the black hole critical behavior also shed light on a
self-similar solution as a critical 
solution (see~\cite{gundlach2003} for a review).
We can say with fair certainty that 
self-similar solutions play important roles near
spacetime singularities. 
Several studies have been done resulting in a complete
classification of self-similar
solutions so far (see~\cite{CarrColey_1} for a review).

Motivated by the above, this paper is intended as the investigation
of particle creation during the naked-singularity formation in self-similar
collapse and of the resulting instability of a CH.
It is shown that irrespective of the details of the model, 
a diverging energy flux is emitted
from a naked shell-focusing singularity forming in generic 
spherically symmetric self-similar spacetime.
The power and energy of particle creation are calculated on the
assumption that the curvature around the singularity causes
particle creation, which was proven by ~\cite{Tanaka}, at least
in the case of self-similar and 
analytic non-self-similar dust models.
Because collapsing matter is not specified, the results
can be applied to several known models of self-similar
collapse. 
Our analysis is regarded as a semiclassical counterpart of~\cite{Nolan},
in which the stability of the CH in self-similar collapse
was tested by a classical field.

This paper is organized as follows. 
In Sec.~\ref{sec:spacetime} a class of self-similar 
spacetimes is introduced and some features of the class of spacetimes
 to be used in later discussion are extracted.
In Sec.~\ref{sec:map}
null geodesic equation is considered to calculate the ``local map'' 
of null rays, which plays a central role in the estimation of 
the power of emissions.
In Sec.~\ref{sec:power} the power and energy
of the created particles are estimated, while redshifts are the focus of
Sec.~\ref{sec:redshift}. Then several underlying
relations among the local map, redshift, and power are found in
Sec.~\ref{sec:relation}. General results obtained in the preceding sections
will be applied to the various models of collapse in
Sec.~\ref{sec:examples}. 
Sec.~\ref{sec:conclusion} is devoted to a summary and a discussion. 
In an Appendix, the same results are derived in terms 
of another coordinate system, which is useful in applications.
The signature $(-,+,+,+)$ and units in which $c=G=\hbar=1$ are used below.

\section{Spherically Symmetric Self-Similar Space-Times Admitting a Naked Singularity}
\label{sec:spacetime}
In this article a class of spacetimes which are spherically symmetric and admitting a homothetic Killing vector field $\xi$, which satisfies $\pounds _{\xi}g_{\mu\nu}=2g_{\mu\nu}$, is considered. The line element of this class of spacetime in an advanced null coordinate system is written as
\begin{eqnarray}
ds^{2}&=&g_{vv}(x)dv^{2}+2g_{vR}(x)dvdR+R^{2}d\Omega^{2},
\label{eq:metric_1}
\end{eqnarray}
where $x\equiv v/R$, $d\Omega^{2}$ is the line element of a unit two
dimensional sphere, and the homothetic Killing vector field is of the form
$\xi=v\partial _{v}+R\partial _{R}$. In this spacetime, the geodesic
equation for an outgoing null ray is
written as
\begin{eqnarray}
\frac{dv}{dR}&=&-\frac{2g_{vR}}{g_{vv}}=xf(x),
\label{eq:dvdR}
\end{eqnarray}
where
\begin{eqnarray}
f(x)\equiv-\frac{2g_{vR}}{xg_{vv}}.
\label{eq:f}
\end{eqnarray}
Equation (\ref{eq:dvdR}) can be written also as
\begin{eqnarray}
\frac{dx}{dR}=\frac{x\left(f(x)-1\right)}{R},
\end{eqnarray}
which is integrated to give
\begin{eqnarray}
\frac{R}{R_{0}}=\exp\left[\int _{x_{0}}^{x}F(x^{\prime})dx^{\prime}\right],\;\;
F(x)\equiv\frac{1}{x\left(f(x)-1\right)},
\label{eq:solution_1}
\end{eqnarray}
where
$x_{0}$ and $R_{0}$ are constants which are related as $R_{0}=R(x=x_{0})$.
The constant $x_{0}$ is chosen as $x_{0}<x^{+}$ and $x_{0}\ne 0$.

\begin{figure}[htbp]
\includegraphics[width=7cm]{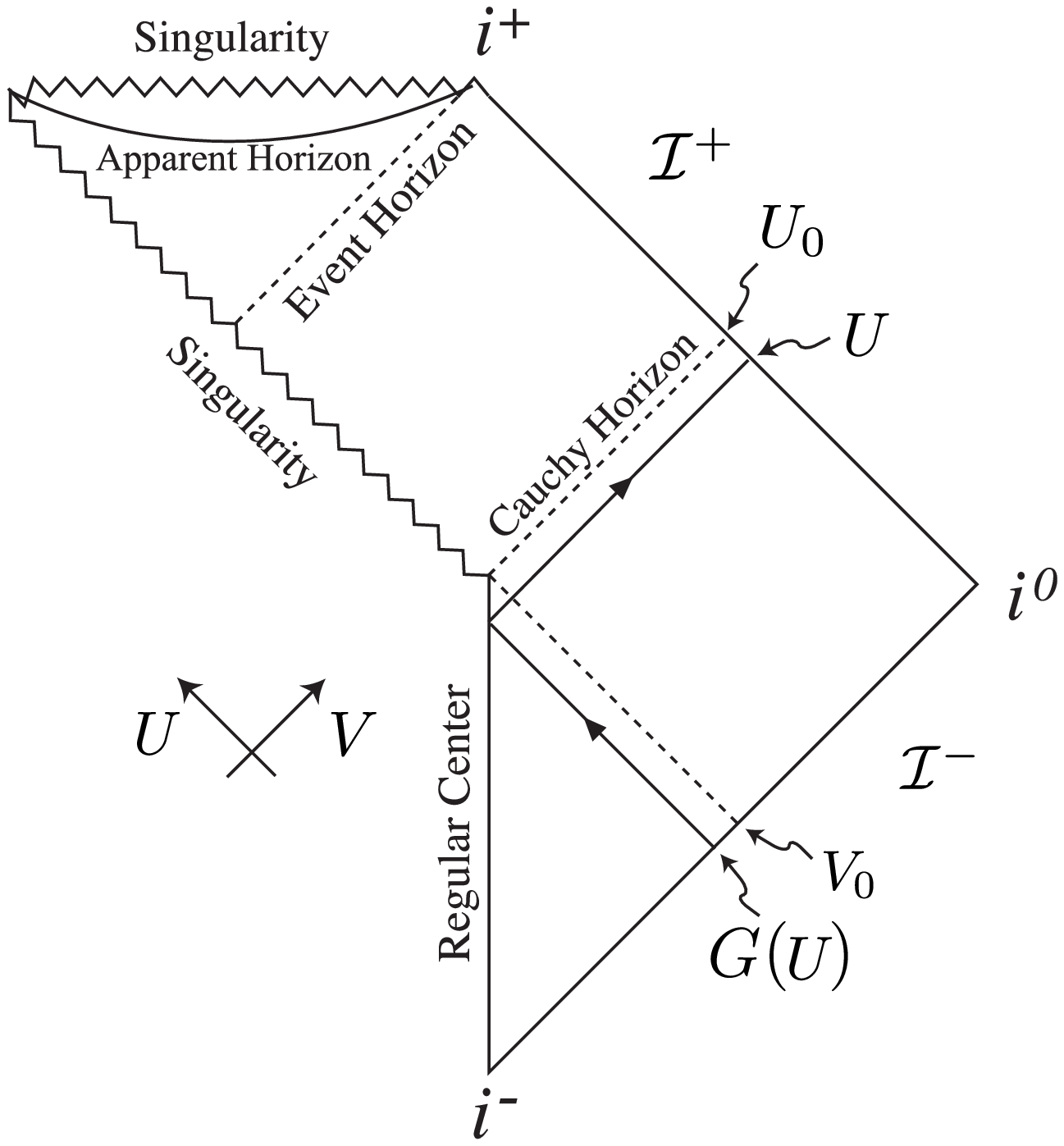}
\caption{\label{fg:penrose}
One of the possible causal structures considered in this article.
A singularity occurs at the spacetime point $(U_{0}, V_{0})$ and is
visible from $\mathcal{I}^{+}$, where $(U, V)$ are suitable double null
coordinates. An outgoing null ray $U=\mathrm{const.}$ can be traced
backward in time from $\mathcal{I}^{+}$ to $\mathcal{I}^{-}$, which turn out to be an ingoing null ray $V=G(U)$. The outgoing null ray $U=U_{0}$ and ingoing null ray $V=V_{0}$ represent the CH and the null ray that terminates at the NS.}
\end{figure}

\begin{figure}[htbp]
\begin{center}
\begin{tabular}{lr}
\includegraphics[width=8cm]{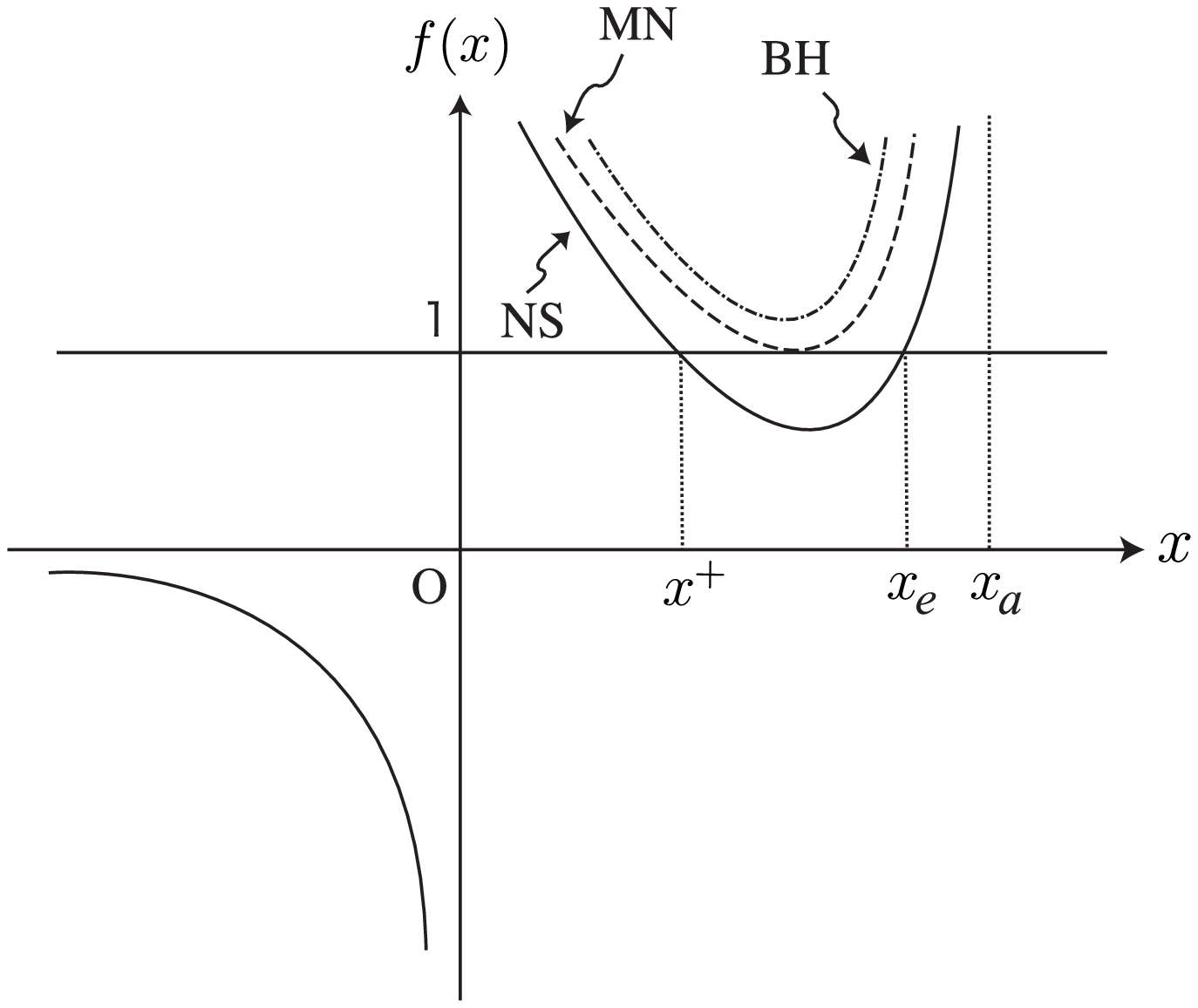} &
\includegraphics[width=6cm]{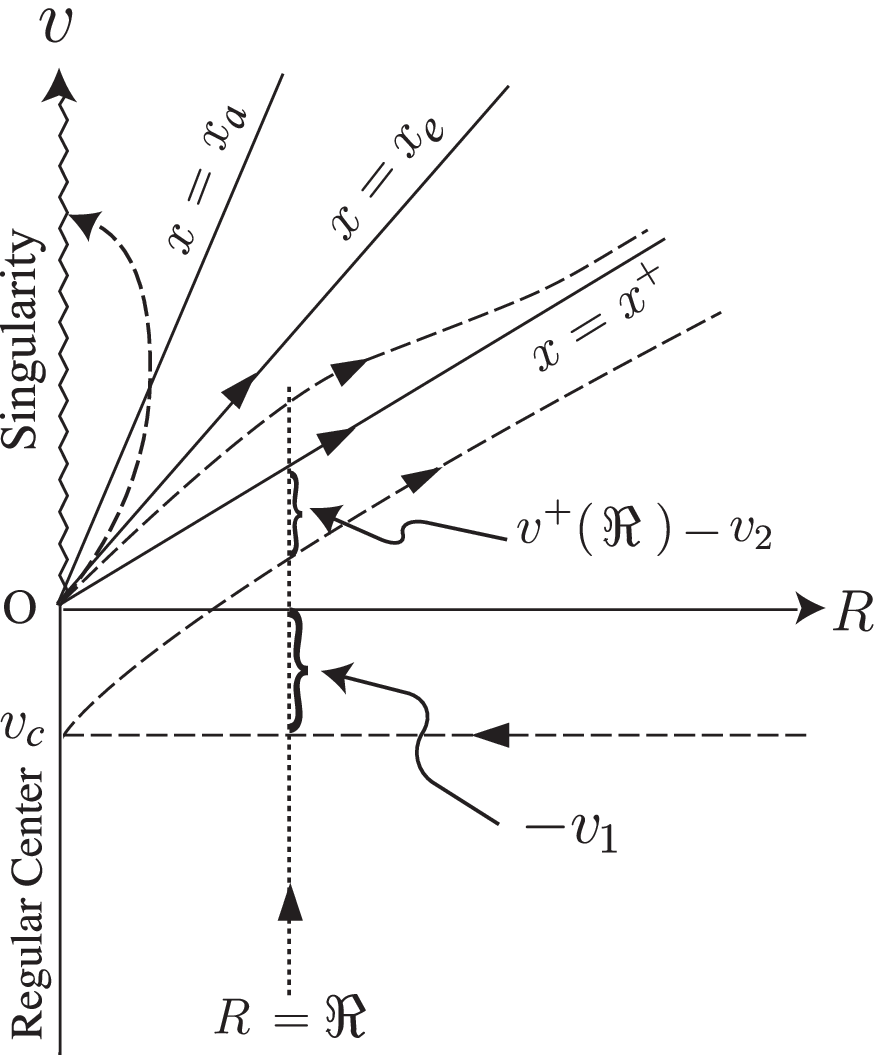}\\
(a) & (b) \\
\end{tabular}
\caption{\label{fg:single_null}
(a) Schematic plots of $f(x)$ defined by Eq.~(\ref{eq:f}) 
for typical collapsing spacetimes which end in a NS or a black
hole. 
Depending on the number of roots of $f(x)=1$, which we denote
by $j$, the causal structure of spacetime changes. The cases of $j=0,\;1,\mbox{and}\;2$ are depicted.
 (i) The case of $j=2$ : $f(x)\,(x>0)$ is depicted by a solid line. The two roots  are
 denoted by $x^{+}$ and
 $x_{e}\,\,(x^{+}<x_{e})$. The geodesics $x=x^{+}$
 and $x=x_{e}$ represent the CH and event horizon,
 respectively. This kind of spacetime admits a NS. 
 (ii) The case of $j=1$ : $f(x)\,(x>0)$ is depicted by a dashed
 line. In this case, $x^{+}=x_{e}$ holds, i.e., the CH and event horizon coincide. This type of singularity is called
 marginally naked (MN). (iii) The case of $j=0$ : $f(x)\,(x>0)$ is depicted
 by a dot-dashed line. In this case the collapse ends in a
 black hole (BH). (b) A typical
spacetime diagram of a collapsing body which ends in a naked
 singularity 
in $(v,R)$ coordinates.
A null ray which is reflected  at the regular center and characteristic
 null rays in respective regions divided by horizons are depicted.
The time intervals $v^{+}(\mathfrak{R})-v_{2}$ and
 $-v_{1}$ in Eq.~(\ref{eq:mapping_1}) are depicted. 
The dotted line is the world line of an observer 
at $R=\mathfrak{R}$.
}  
\end{center}
\end{figure}

What we have to do first is to extract features of $f(x)$, which determines
the spacetime structure.
The Misner-Sharp mass in this spacetime is given by
\begin{eqnarray*}
m(v, R)\equiv\frac{R}{2}\left(1-\nabla _{\mu}R\nabla^{\mu}R\right)
=\frac{R}{2}\left(1+\frac{4}{x^{2}f^{2}g_{vv}}\right).
\end{eqnarray*}
The regularity of the center 
$R=0$ in the region $v<0$ and the
absence of a trapped or a marginally trapped surface 
for $0<R$ and $v\leq 0$ are assumed.
The latter condition is $\nabla _{\mu}R\nabla^{\mu}R>0$ for all $x\in (-\infty, 0]$,
which is written as $g_{vv}<0$ for all $x\in (-\infty, 0]$ in the present case.
The inevitability of a curvature singularity at the origin
$v=R=0$ can be shown except for a flat spacetime~\cite{Nolan}.
In this article we consider self-similar spacetimes with a globally naked singularity, of which existence breaks the weak version of the cosmic censorship hypothesis.
One of the possible causal structures of the naked-singular spacetimes is depicted
in Fig.~\ref{fg:penrose}.
The coordinate $v$ is set to be the proper time along the regular center to remove a gauge freedom $v\to V(v)$, so that $\lim _{x\to-\infty}g_{vv}=-1$. When $m/R^{3}$ is required 
to be finite at the regular center, the function $f(x)$ behaves as 
\begin{eqnarray}
f\simeq 2/x \;\;\mbox{as}\;\;x\to -\infty.
\label{eq:condition_1}
\end{eqnarray}
The quantity $m/R^{3}$ must be finite also in the limit $v\to 0$ for fixed $R\;(>0)$ so that
\begin{eqnarray}
f=O(|x|^{\beta}) \;\;\mbox{as}\;\;x\to 0,
\label{eq:condition_2}
\end{eqnarray}
where $\beta\leq-1$.
When there are positive roots of the algebraic equation $f(x)=1$, 
it can be shown that the curve $x=x^{+}$ is a CH,
as we will see in Appendix~\ref{sec:CH},
where $x^{+}$ is the smallest root.
The differentiability of the metric function $f$ is assumed to be as follows:
\begin{eqnarray}
\frac{1}{f} \in C^{0}\left((-\infty, x^{+})\right),\;\;
f\in C^{2-}\;\;\mbox{at $x=x^{+}$.}
\label{eq:condition_3}
\end{eqnarray}
The former condition guarantees the existence and uniqueness of geodesics in this system. It is also  assumed~\cite{footnote_2} that
\begin{eqnarray}
f^{\prime}(x^{+})<0.
\label{eq:condition_4}
\end{eqnarray}
The schematic plot of the function $f(x)$ is shown in Fig.~\ref{fg:single_null}(a).


\section{Local Map}
\label{sec:map}
To estimate the power of particle creation just before the singularity occurs,
the pole at $x=x^{+}$ should be
extracted from the integrand in Eq.~(\ref{eq:solution_1}) as follows:
\begin{eqnarray}
\frac{R}{R_{0}}&=&\exp\left[\int _{x_{0}}^{x}\left\{F(x^{\prime})-\frac{1}{\gamma(x^{\prime}-x^{+})}\right\}\right]\exp\left[\int _{x_{0}}^{x}\frac{dx^{\prime}}{\gamma(x^{\prime}-x^{+})}\right]\nonumber\\
&=&\exp\left[\int _{x_{0}}^{x}F^{\ast}(x^{\prime})dx^{\prime}\right]\left(\frac{x^{+}-x}{x^{+}-x_{0}}\right)^{1/\gamma},
\label{eq:leading_1}
\end{eqnarray}
where
\begin{eqnarray}
\gamma&\equiv&x^{+}f^{\prime}(x^{+}),\label{eq:gamma}\\
F^{\ast}(x)&\equiv&F(x)-\frac{1}{\gamma(x-x^{+})}.\nonumber
\end{eqnarray}

The constant $R_{0}$ in Eq.~(\ref{eq:solution_1}), which parameterizes solutions, 
is related to the time $v_{c}\equiv v(R=0)<0$
when the outgoing null ray emanates from the regular center as follows:
\begin{eqnarray}
\frac{R}{R_{0}}&=&\exp\left[-\int_{x_{0}}^{x}\frac{dx^{\prime}}{x^{\prime}}\right]\exp\left[\int _{x_{0}}^{x}\left\{F(x^{\prime})+\frac{1}{x^{\prime}}\right\}dx^{\prime}\right]\nonumber\\
&=&\left|\frac{Rx_{0}}{v}\right|\exp\left[\int _{x_{0}}^{x}\frac{f(x^{\prime})}{x^{\prime}\left(f(x^{\prime})-1\right)}dx^{\prime}\right].
\label{eq:asymptotic_1}\nonumber
\end{eqnarray}
Taking the limit of $R\to 0$ $(v<0)$, following relation is obtained:
\begin{eqnarray}
R_{0}=-\frac{v_{c}}{|x_{0}|I},\;\;
I\equiv\exp\left[\int _{x_{0}}^{-\infty}\frac{f(x^{\prime})}
{x^{\prime}\left(f(x^{\prime})-1\right)}dx^{\prime}\right],
\label{eq:initial_1}
\end{eqnarray}
where condition~(\ref{eq:condition_1}) ensures the convergence of the integral in Eq.~(\ref{eq:initial_1}).

Combination of Eqs.~(\ref{eq:leading_1}) and (\ref{eq:initial_1})
yields
\begin{eqnarray}
R=C(R, x)\left(v^{+}(R)-v\right)^{1/\gamma}v_{c},
\label{eq:radial_1}
\end{eqnarray}
where 
\begin{eqnarray}
v^{+}(R)&\equiv&x^{+}R,\nonumber\\
C(R, x)&\equiv&-|x_{0}|^{-1}I^{-1}\left[(x^{+}-x_{0})R\right]^{-1/\gamma}\exp\left[\int _{x_{0}}^{x}F^{\ast}(x^{\prime})dx^{\prime}\right].\nonumber
\label{eq:C_1}
\end{eqnarray}
Before turning to the derivation of the local map,
a few remarks should be made concerning the function $C(R, x)$.
Due to condition (\ref{eq:condition_3}), 
$C(R, x)$ converges to some finite constant in the limit
$x\to x^{+}$ for fixed $R$. 
The dependence of $C(R, x)$ on $x_{0}$ is only an apparent one as
\begin{eqnarray}
\frac{\partial C}{\partial x_{0}}=0,
\label{eq:Derivative_C_1}
\end{eqnarray} 
which we use in Sec.~\ref{sec:relation}.

Now let us consider a pair of ingoing and outgoing null rays
such that the latter is the reflection of the former at the regular
center. 
An observer who rests at $R=\mathfrak{R}$ will
encounter the null ray twice so that we denote the time of first encounter
by $v_{1}$ and that of the second by $v_{2}$.
The relation between $v_{1}$ and $v_{2}$, which we call the local map,
is obtained from Eq.~(\ref{eq:radial_1}):
\begin{eqnarray}
v_{1}=\frac{\mathfrak{R}}{C(\mathfrak{R}, x_{2})}\left(v^{+}(\mathfrak{R})-v_{2}\right)^{\alpha _{1}},\label{eq:mapping_1}
\end{eqnarray}
where 
\begin{eqnarray}
x_{2}&\equiv& v_{2}/\mathfrak{R},\nonumber\\
\alpha _{1}&\equiv&-1/\gamma. 
\label{eq:alpha_null}
\end{eqnarray}
The time intervals $-v_{1}\;(>0)$
and $v^{+}(\mathfrak{R})-v_{2}$ are 
depicted schematically in Fig.~\ref{fg:single_null}(b). 
It is noted that the result does not change even if we choose a small value of $\mathfrak{R}$.
Namely, the nature of the local map is determined by the behavior of null rays near the singularity.

\section{Power and Energy}
\label{sec:power}
A global map $V=G(U)$ is defined to be a relation
between the moments when one null ray leaves $\mathcal{I}^{-}$
and terminates at $\mathcal{I}^{+}$ after passing through the regular center
(see Fig.~\ref{fg:penrose}).
Assuming the geometric optics approximation, one can obtain the power of emission as the vacuum expectation value of a stress-energy tensor by the point-splitting regularization from the global map~\cite{Ford},
\begin{eqnarray*}
P=\frac{1}{24\pi}\left[\frac{3}{2}\left(
\frac{G^{\prime\prime}}{G^{\prime}}\right)^2-
\frac{G^{\prime\prime\prime}}{G^{\prime}}\right]
\end{eqnarray*}
for a minimally coupled scalar field, and 
\begin{eqnarray*}
\hat{P}=\frac{1}{48\pi}\left(\frac{G^{\prime\prime}}{G^{\prime}}\right)^{2}
\end{eqnarray*}
for a conformally coupled scalar field.

Spherically symmetric self-similar spacetimes are not asymptotically
flat in general. 
They therefore should be matched with an outer asymptotically flat region
via a proper non-self-similar region.
This matching procedure is quite straightforward
for dust collapse~\cite{BarveBarve2}.
Although it seems to be necessary to solve null geodesic equations
in such a ``patched-up'' spacetime, the main properties of the global
map must be determined by the
behavior of null rays passing near the point where the singularity
occurs. This expectation has been confirmed in \cite{Tanaka}, at least
for the self-similar and analytic dust models. Therefore, one can safely assume that
the global map inherits the main properties of the local map, such as the
value of the exponent and differentiability. This means that from
Eq.~(\ref{eq:mapping_1}), the asymptotic form
of the global map would take the form
\begin{eqnarray}
G(U)&=&V_{0}-\left(U_{0}-U\right)^{\alpha}G_{\ast}(U),\nonumber
\end{eqnarray}
where the null rays $U=U_{0}$ and $V=V_{0}$ are the CH and the ingoing null ray that
terminates at the NS, respectively. The function $G_{\ast}(U)$ is a regular function which does not vanish at the CH and $\alpha$ is the exponent of the local map,
$\alpha _{1}$ in Eq.~(\ref{eq:mapping_1}).

In the case of $\alpha=1$, the leading contribution to the power of particle creation
is calculated as
\begin{eqnarray}
P&=&\frac{2G_{\ast}^{\prime 2}(U_{0})-G_{\ast}(U_{0})G_{\ast}^{\prime\prime}(U_{0})}{8\pi G_{\ast}^{2}(U_{0})},\label{eq:power_3}\\
\hat{P}&=&\frac{1}{12\pi}\left(\frac{G_{\ast}^{\prime}(U_{0})}{G_{\ast}(U_{0})}\right)^{2}
,\label{eq:power_4}
\end{eqnarray}
so that the power remains finite at the CH.
Unfortunately $G_{\ast}(U_{0})$ and its derivatives cannot be known until the null geodesic equation is solved globally, so that one could not know the power of emission from only the information contained in the local map. In terms of redshift, $\alpha=1$ corresponds to the case that the redshift of a particle remains finite at the CH, as we will see in Sec.~\ref{sec:redshift}.

On the other hand,
in the case of $\alpha\neq 1$ the leading contribution 
is obtained as
\begin{eqnarray}
P=\frac{\alpha^{2}-1}{48\pi\left(U_{0}-U\right)^{2}},\label{eq:power_1}
\end{eqnarray}
for a minimally coupled scalar field. For a conformally coupled one, the power of emission
is obtained by replacing the factor $(\alpha^{2}-1)$ in Eq.~(\ref{eq:power_1}) for
$(\alpha-1)^{2}$. 
The power is proportional to the inverse square of the remaining time to the CH.
If $\alpha>1$, the power diverges to positive infinity for both minimally and conformally
coupled scalar fields,
while if $0<\alpha<1$, the power diverges to negative and positive infinity
for minimally and conformally coupled scalar fields, respectively.
In terms of the redshift of particles, the case that $\alpha>1$ ($0<\alpha<1$)
corresponds to infinite redshift (blueshift) at the CH,
as we will see in Sec.~\ref{sec:redshift}.
The emitted energy can be estimated as
\begin{eqnarray}
E=\int _{-\infty}^{U}P(U^{\prime})dU^{\prime}
=\frac{\alpha^{2}-1}{48\pi\left(U_{0}-U\right)}.
\end{eqnarray}
Although the emitted energy diverges when the CH is approached, this divergence needs to be regarded carefully. The semiclassical approximation would cease to be valid when the curvature radius at some spacetime point inside star reaches the Planck scale.
Here we make a natural assumption that such a situation happens at the center of a star at $v=-t_{\mathrm{QG}}$~\cite{footnote_3}. In the case of $\alpha>1$, it can be expected that for a ray emanating from the center at $v=-t_{\mathrm{QG}}$, the time difference $U_{0}-U$ would be greater than the order of $t_{\mathrm{QG}}$ due to redshift, i.e., $\Delta U\equiv U_{0}-U>t_{\mathrm{QG}}$. Then energy
emitted by the time $U_{0}-\Delta U$ is
\begin{eqnarray}
E=\frac{\alpha^{2}-1}{48\pi\Delta U}
<\frac{\alpha^{2}-1}{48\pi}E_{\mathrm{QG}},
\end{eqnarray}
where $E_{\mathrm{QG}}\equiv 1/t_{\mathrm{QG}}$.
If the factor $\left(\alpha^{2}-1\right)/(48\pi)$ is on the order of unity,
the total radiated energy within the semiclassical
phase is less than the order of $E_{\mathrm{QG}}$,
which would be of course much less than the mass of ordinary astrophysical stars.
It would be better to say that a collapsing star would enter the phase of
quantum gravity with most of its mass intact.
Therefore, one could not predict whether a star which collapses
to a NS evaporates away or ceases to radiate at its final epoch.
This has been pointed out in \cite{NSEandQG} after careful investigation.
We should not overlook that this feature is much different from that of black hole evaporation,
in which quantum gravitational effects appear after a black hole loses almost all its mass.

One can recognize, however,
that if $\alpha\gg 1$ the radiated energy could be large.
This situation is realized in the limit to marginally NS,
in which the CH and event horizon coincide~\cite{footnote_4}.
To illustrate the unbounded increase of $\alpha$ in this limit,
we have to look deeper into the causal structure, which is determined by the function $f$.
We order the positive roots of the equation $f(x)=1$ as $0<x_{1}=x^{+}<x_{2}<\cdot\cdot\cdot<x_{n}$, where we count multiple roots as one root. The existence of $x_{a}$ satisfying 
$\lim _{x\to x_{a}}f(x)=+\infty$ 
with $x_{n}<x_{a}$ and the continuity of $f$ in the region
$x_{1}<x<x_{a}$ are assumed.
In the region $x\in\left(x_{n}, x_{a}\right)$,
$dR/dv>0$ along the null geodesics and
$\lim _{x\to x_{a}}dR/dv=0$ from Eq.~(\ref{eq:dvdR}). This implies that
outgoing null rays in this region are to turn back in
the direction of the singularity at $x=x_{a}$ and that
the curve $x=x_{n}$ is the last outgoing null ray which
can escape to infinity. That is to say, $x=x_{a}$ and
$x=x_{n}\equiv x_{e}$ are the apparent horizon and event horizon, respectively. 
Hereafter, we shall concentrate on the case of $n=2$. The function $f(x)$ would be written as
\begin{eqnarray}
f(x)-1=f_{\ast}(x)(x-x^{+})\left(x_{e}-x\right)^{m},\;\;0<x<x_{a},
\label{eq:f-1}
\end{eqnarray}
where $f_{\ast}(x)$ is a function which satisfies $f_{\ast}(x^{+})<0$ and $m$ is some positive integer. The exponent of the factor $(x-x^{+})$ is restricted to unity because of the condition $f^{\prime}(x^{+})<0$ and the differentiability $f\in C^{2-}$ at $x=x^{+}$.
With Eq.~(\ref{eq:f-1}), the exponent of the local map is calculated as
\begin{eqnarray}
\alpha _{1}\equiv -\frac{1}{x^{+}f^{\prime}(x^{+})}=-\frac{1}{x^{+}f_{\ast}^{\prime}(x^{+})\left(x_{e}-x^{+}\right)^{m}},
\end{eqnarray}
to show that $\alpha _{1}$ can be arbitrarily large in the limit
$x^{+}\to x_{e}$.


\section{Redshift}
\label{sec:redshift}
The estimation of redshift of the radial null ray
would help us understand the behavior of the power and would be necessary
for discussing the validity of geometric optics and semiclassical
approximations. 
Hereafter the tangent vector of the null ray is denoted by $k^{\mu}\equiv dx^{\mu}/d\lambda$, where $\lambda$ is an affine parameter.

With the null condition $k^{\mu}k_{\mu}=0$, the $v$-component of the geodesic equation $k^{\mu}\nabla _{\mu}k^{\nu}=0$ leads to  
\begin{eqnarray}
\frac{dk^{v}}{d\lambda}+\frac{(k^{v})^{2}}{R}\left(\frac{1}{g_{vR}}\frac{dg_{vR}}{dx}+\frac{x}{2}\frac{1}{g_{vR}}\frac{dg_{vv}}{dx}\right)=0.\nonumber
\end{eqnarray}
Furthermore by using the relation
\begin{eqnarray*}
\frac{d}{d\lambda}=\frac{k^{v}}{R}\left(1+\frac{xg_{vv}}{2g_{vR}}\right)\frac{d}{dx},
\end{eqnarray*}
$k^{v}(x)$ is integrated to give
\begin{eqnarray}
\frac{k^{v}(x)}{k^{v}_{0}}=\exp\left[\int^{x}_{\tilde{x}_{0}}\tilde{F}(x^{\prime})dx^{\prime}\right]=\exp\left[\int _{\tilde{x}_{0}}^{x}\tilde{F}^{\ast}(x^{\prime})dx^{\prime}\right]\left(\frac{x^{+}-x}{x^{+}-\tilde{x}_{0}}\right)^{-(1+\gamma)/\gamma},
\label{eq:frequency_1}
\end{eqnarray}
where 
\begin{eqnarray*}
\tilde{F}(x)&\equiv&-\frac{1}{g_{vR}}\frac{dg_{vR}}{dx}+\frac{1}{1-f}\left(\frac{1}{x}+\frac{1}{f}\frac{df}{dx}\right),\\
\tilde{F}^{\ast}(x)&\equiv& \tilde{F}(x)+\frac{1+\gamma}{\gamma}\frac{1}{x-x^{+}}.
\end{eqnarray*}
The constant $\tilde{x}_{0}$, which is set as $\tilde{x}_{0}<x^{+}$, and the constant $k^{v}_{0}$ are
related as $k^{v}_{0}=k^{v}(\tilde{x}_{0})$.

The constant $k^{v}_{0}$ is related to $k^{v}_{c}\equiv k^{v}(R=0)$ as
\begin{eqnarray}
k^{v}_{0}=\frac{k^{v}_{c}}{\tilde{I}},\;\;
\tilde{I}\equiv\exp\left[\int^{-\infty}_{\tilde{x}_{0}}\tilde{F}(x^{\prime})dx^{\prime}\right].
\label{eq:initial_k1}
\end{eqnarray}
Combination of Eqs.~(\ref{eq:frequency_1}) and (\ref{eq:initial_k1})
yields 
\begin{equation}
k^{v}(x)=\tilde{C}(R,x)\left(v^{+}(R)-v\right)^{\alpha _{1}-1}k^{v}_{c},
\label{eq:total_redshift_1}
\end{equation}
where 
\begin{eqnarray*}
\tilde{C}(R,x)\equiv\tilde{I}^{-1}\left[(x^{+}-\tilde{x}_{0})R\right]^{(1+\gamma)/\gamma}\exp\left[\int _{\tilde{x}_{0}}^{x}\tilde{F}^{\ast}(x^{\prime})dx^{\prime}\right].
\end{eqnarray*}

Now, let us consider time-like observers who rest at $R=0$
and 
$R=\mathfrak{R}\;(d\theta=d\phi=0)$.
The observed frequency is given by
$\hat{\omega}\equiv-u_{\mu}k^{\mu}=\sqrt{|g_{vv}(x)|}k^{v}(x)$, where $u_{\mu}$ is the four-velocity of observer.
When $\hat{\omega} _{1}\equiv\lim _{x\to-\infty}\hat{\omega}(x)$ and $\hat{\omega} _{2}\equiv\lim _{x\to x^{+}}\hat{\omega}(x)$ are defined, Eq.~(\ref{eq:total_redshift_1}) yields
\begin{eqnarray}
\frac{\hat{\omega} _{2}}{\hat{\omega}_{1}}=\sqrt{\left|\frac{g_{vv}(x^{+})}{g_{vv}(-\infty)}\right|}\tilde{C}(\mathfrak{R}, x_{2})(v^{+}(\mathfrak{R})-v_{2})^{\alpha _{1}-1}.
\label{eq:omega_1}
\end{eqnarray} 
Thus we see that if $\alpha_ {1}>1$ ($0<\alpha _{1}<1$) the redshift (blueshift) of emitted particle diverges at the CH, while it remains finite if $\alpha _{1}=1$.
The relation between the redshift derived above and the local map will be
presented in the next section.

\section{Relations among the Local Map, Power, and Redshift}
\label{sec:relation}
There would be a relation between the local map and redshift because the local map 
describes a kind of time delay.
Since the asymptotic behavior of the local map and
redshift in the limit $x\to x^{+}$ is considered here,
the time dependence is omitted as $C(R,x)\to C(R)$. From Eq.~(\ref{eq:mapping_1}), the relation
\begin{eqnarray}
\frac{dv_{2}}{dv_{1}}=\frac{\gamma C(\mathfrak{R})}{\mathfrak{R}}(v^{+}(\mathfrak{R})-v_{2})^{(1+\gamma)/\gamma}
\label{eq:derivative}
\end{eqnarray}
holds to give an alternative definition of redshift.
Indeed, the time dependence in Eq.~(\ref{eq:derivative}) can be replaced with the ratio of $k^{v}$ by Eq.~(\ref{eq:total_redshift_1}) as 
\begin{equation}
\frac{dv_{2}}{dv_{1}}
=\left|\frac{g_{vv}(-\infty)}{g_{vv}(x^{+})}\right|\frac{k^{v}_{c}}{k^{v}(x_{2})},
\label{eq:relation_1}
\end{equation}
where we set $x_{0}=\tilde {x}_{0}$ in the evaluation of the integral in $C$ to derive Eq.~(\ref{eq:relation_1}) since $C$ does not depend on $x_{0}$ from Eq.~(\ref{eq:Derivative_C_1}).
Equation~(\ref{eq:relation_1}) can be written as
\begin{eqnarray}
\frac{d\tau _{2}}{d\tau _{1}}=\frac{\hat{\omega}_{1}}{\hat{\omega}_{2}},
\label{eq:relation_2}
\end{eqnarray}
where $d\tau _{i}\equiv\sqrt{|g_{vv}|}dv_{i}$ $(i=1,2)$ is the proper time
measured by the observer.
This relates the time delay and redshift
to reveal that the redshift essentially
corresponds to the local map and also to confirm the consistency
of the analyses in Secs. \ref{sec:map} and \ref{sec:redshift}.

There exists a plausible relation also between the power of emission and the
redshift of particles as mentioned in Sec.~\ref{sec:power}.
In the case of $\alpha>1$ ($0<\alpha<1$),
the power and redshift (blueshift)
diverge at the CH from Eqs.~(\ref{eq:power_1}) and (\ref{eq:omega_1}),
while the power and redshift
remain finite at the CH in the case of $\alpha=1$.
We may, therefore, reasonably conclude
that the divergence of the redshift or blueshift 
at the CH causes that of the power.

\section{Examples}
In this section,
we will take examples to illustrate how the features
obtained in the previous sections are realized in concrete
models.
Since several models are written in the diagonal form of
a metric tensor, the formulation and notations 
for the diagonal form of a metric tensor are developed in Appendix \ref{sec:Diagonal}.
\label{sec:examples}
\subsection{Minkowski spacetime}
\label{subsec:minkowski_case}
Although neither singularity nor horizon exists
in Minkowski spacetime,
one can test the formalism by applying it to this trivial spacetime.
The line element is written as
\begin{eqnarray*}
ds^{2}=-dv^{2}+2dvdR+R^{2}d\Omega^{2},
\end{eqnarray*}
or
\begin{eqnarray*}
ds^{2}=-dt^{2}+dr^{2}+r^{2}d\Omega^{2}.
\end{eqnarray*}
From definitions~(\ref{eq:f}) and (\ref{eq:wpm}), one obtains
\begin{eqnarray*}
f(x)=\frac{2}{x},\;\;
w_{\pm}(z)=\pm\frac{1}{z}.
\end{eqnarray*}
The roots of the algebraic equations $f(x)=1$ and $w_{\pm}(z)=1$ are
given as $x^{+}=2$ and $z^{\pm}=\pm 1$, respectively. 
It must be noted that the function $f(x)$
satisfies conditions (\ref{eq:condition_1})-(\ref{eq:condition_4}),
which have been assumed to derive the local map.
According to Eqs.~(\ref{eq:gamma}),
 (\ref{eq:alpha_null}), (\ref{eq:gamma_pm}), and
(\ref{eq:alpha_diagonal}), one can easily check that the exponents of local
maps, $\alpha _{1}$ and $\alpha _{2}$, are
unity. This fact just tells us that the
redshift and power remain finite in a flat spacetime.

\subsection{Vaidya Solution}
\label{subsec:vaidya_case}
The Vaidya solution describes the collapse
of a null-dust fluid~\cite{Vaidya_1Vaidya_2Vaidya_3}.
The global map for the self-similar Vaidya collapse was derived in \cite{Singh}. 

The line element in the self-similar Vaidya spacetime is written as
\begin{eqnarray}
ds^{2}=-\left(1-m(x)\right)dv^{2}+2dvdR+R^{2}d\Omega^{2},\nonumber
\end{eqnarray}
where $m(x)=0$ for $x<0$ and $m(x)=2\mu x$ for $x\geq 0$.
The constant $\mu$ is restricted as $0<\mu<1/16$
for the nakedness, so that the spacetime with $\mu=1/16$
corresponds to a marginally naked-singular one.
The function $f(x)$ is written as
\begin{eqnarray}
f(x)=\frac{2}{x(1-m(x))}.\nonumber
\end{eqnarray}
The roots of algebraic equation $f(x)=1$ are given as
\begin{eqnarray*}
x^{+}=\frac{1-\sqrt{1-16\mu}}{4\mu}\quad\mbox{and}\quad
x_{e}=\frac{1+\sqrt{1-16\mu}}{4\mu}.
\end{eqnarray*}
What has to be noticed is that $f$ satisfies conditions (\ref{eq:condition_1})-(\ref{eq:condition_4}).
The exponent is calculated as
\begin{eqnarray*}
\alpha _{1}\equiv -\frac{1}{x^{+}f^{\prime}(x^{+})}
=\frac{1+\sqrt{1-16\mu}}{2\sqrt{1-16\mu}}.
\end{eqnarray*}
This exponent coincides with that of the global map which was
obtained in \cite{Singh}.
One can see that $\lim _{\mu\to 0}\alpha _{1}=1$ and $\lim _{\mu\to
1/16}\alpha _{1}=\infty$, where the former and latter correspond
to the limits of Minkowski and marginally naked-singular spacetime.
This example is a good illustration of the efficiency of the local map
and the divergence of $\alpha _{1}$ in the limit to marginally NS.

\subsection{Roberts Solution}
\label{subsec:roberts_case}
\begin{figure}[htbp]
\includegraphics[width=4.5cm]{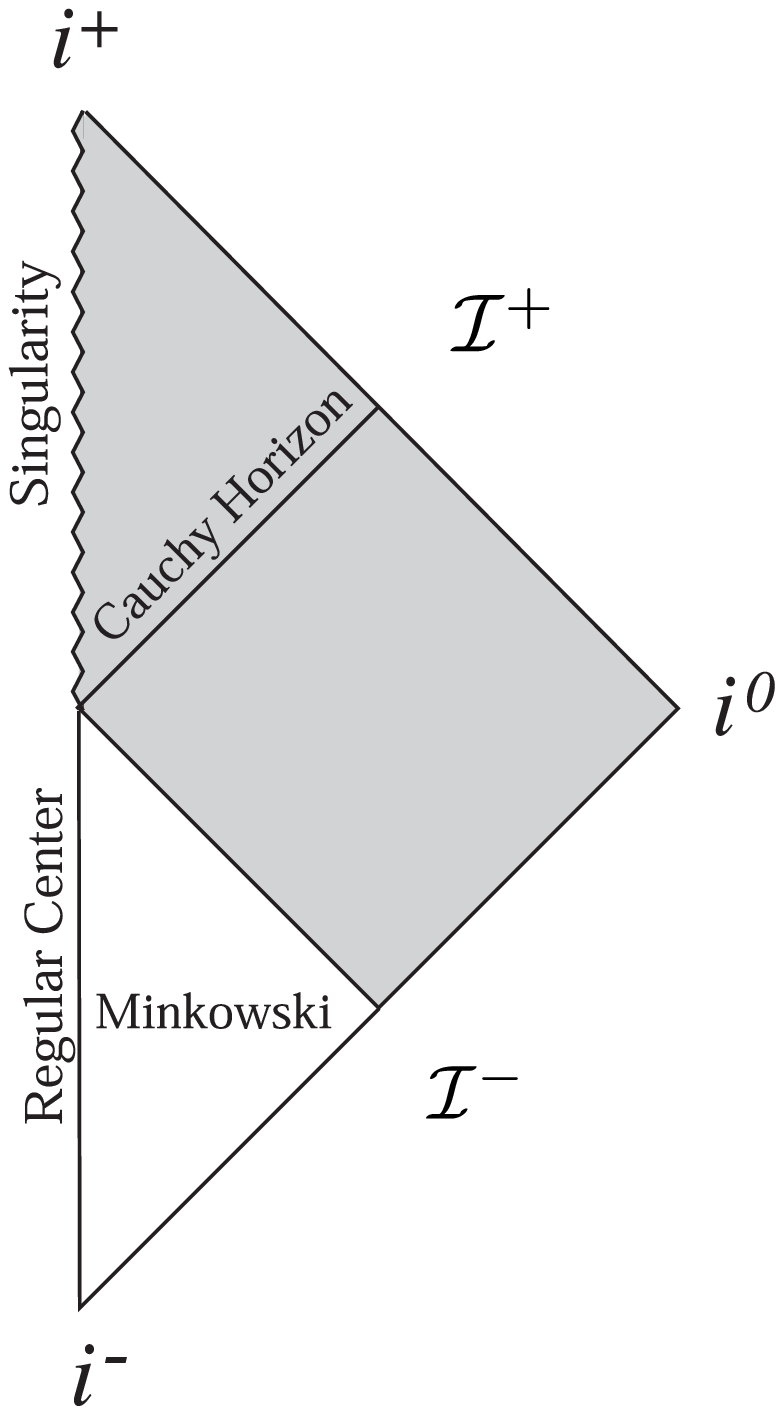}
\caption{\label{fg:roberts}
The conformal diagram of the Roberts solution Eq.~(\ref{eq:roberts}) with $|\sigma|<1/2$. The region $v<0$ is flat and the shading region is filled with a collapsing massless scalar field. A time-like NS occurs at $v=0$.}  
\end{figure}

The Roberts solution describes the self-similar collapse of
a massless scalar field~\cite{Roberts}. 
The line element is given as 
\begin{eqnarray}
ds^{2}=-\left(1-\frac{2h(x)h^{\prime}(x)}{\sqrt{1+h^{2}(x)}}\right)dv^{2}+\frac{2}{\sqrt{1+h^{2}(x)}}dvdR+R^{2}d\Omega^{2},\label{eq:roberts}
\end{eqnarray}
where $h(x)=0$ for $x<0$ and $h(x)=\sigma x$ for $x \geq 0$,
so that the region with negative $v$ is flat.
The constant $\sigma$ is restricted here as $|\sigma|<1/2$ for the causal structure such that the spacetime has a time-like NS as in Fig.~\ref{fg:roberts}. The function $f(x)$ is written as
\begin{eqnarray}
f(x)=\frac{2}{x(\sqrt{1+h^{2}}-2hh^{\prime})}.\nonumber
\end{eqnarray}
The conditions on $f(x)$ are again satisfied.
The algebraic equation $f(x)=1$ has a positive root $x^{+}=2/\sqrt{1-4\sigma^{2}}$ so that
the exponent of local map is calculated as
\begin{eqnarray*}
\alpha _{1}\equiv -\frac{1}{x^{+}f^{\prime}(x^{+})}=1
\end{eqnarray*} 
for $|\sigma|<1/2$.
Thus we see that the Roberts solution provides a non-trivial example of spacetime in which the power of particle creation remains finite at the CH in self-similar collapse.
\subsection{Lema\^{\i}tre-Tolman-Bondi Solution}
\label{sec:ltb_case}
The Lema\^{\i}tre-Tolman-Bondi (LTB) solution describes the collapse of a dust fluid~\cite{LemaitreTolmanBondi}. Although both global and local maps were derived in \cite{BarveBarve2} and \cite{Tanaka} respectively for self-similar LTB collapse, the exponent in \cite{BarveBarve2}  is reproduced from the formalism in Sec.~\ref{sec:map}. 
 
The line element of self-similar LTB spacetime (for example, see~\cite{BarveBarve2}) is
\begin{eqnarray}
ds^{2}=-dt^{2}+\left[\frac{1-az/3}{(1-az)^{1/3}}\right]^{2}dr^{2}+r^{2}(1-az)^{4/3}d\Omega^{2},\nonumber
\end{eqnarray}
where the constant $a$ is related to a ``mass parameter'' $\lambda$ as
$a=\frac{3}{2}\sqrt{\lambda}$. The constant $\lambda$ is restricted to
the range $0<\lambda<6(26-15\sqrt{3})\equiv \lambda _{m}$,
where the latter inequality is imposed by the nakedness of the
singularity~\cite{Joshi}, so that the spacetime with $\lambda=\lambda
_{m}$ corresponds to marginally naked-singular one.
One obtains the function $w_{\pm}(z)$ according to definition~(\ref{eq:wpm}) as
\begin{eqnarray}
w_{\pm}(z)=\pm\frac{1-az/3}{z\left(1-az\right)^{1/3}}.
\label{eq:wpm_LTB}
\end{eqnarray}
The required conditions on $w_{\pm}$ in calculating 
the local map are satisfied.
When a new variable $y\equiv\left(1-az\right)^{1/3}$ introduced, Eq.~(\ref{eq:wpm_LTB})
can be written as
\begin{eqnarray}
w_{\pm}^{2}(z)=1-\frac{4g_{+}(y)g_{-}(y)}{\left[g_{+}(y)+g_{-}(y)\right]^{2}},\nonumber
\end{eqnarray}
where $g_{\pm}(y)\equiv 3y^{4}\mp ay^{3}-3y\mp 2a$, so that the roots of
algebraic equations $w_{\pm}(z)=1$ correspond to those of
$g_{\mp}(y)=0$. Using the chain rule $d/dz=(dy/dz)d/dy$, one obtains
\begin{eqnarray}
z^{\pm}w_{\pm}^{\prime}(z^{\pm})=\frac{2(1-3\alpha _{\mp}^{3})g^{\prime}_{\mp}(\alpha_{\mp})}{3\alpha _{\pm}g_{\pm}(\alpha _{\mp})},\nonumber
\end{eqnarray}
where $\alpha _{\pm}\equiv\left(1-az^{\mp}\right)^{1/3}$ and the prime denotes differentiation with respect to the argument of the function.
The exponent of local map is calculated as
\begin{eqnarray*}
\alpha _{2}\equiv\frac{z^{-}w_{-}^{\prime}(z^{-})}{z^{+}w_{+}^{\prime}(z^{+})}
=\frac{\alpha _{-}^{3}g^{\prime}_{+}(\alpha _{+})}{\alpha _{+}^{3}g^{\prime}_{-}(\alpha _{-})}
\end{eqnarray*}
to coincide with the one derived
in~\cite{BarveBarve2,Tanaka}. This exponent becomes close to unity as
$\lambda\to 0$ and increases
monotonically with $\lambda$ to diverge to infinity as 
$\lambda\to\lambda _{m}$.
\subsection{General Relativistic Larson-Penston Solution}
\label{subsec:grlp_case}
As the last example the general relativistic Larson-Penston
(GRLP) solution, which describes the self-similar collapse
 of a perfect fluid~\cite{OriPiran3,OriPiranOriPiran2}, is considered.
For the present, it may be useful to review the GRLP
solution and its importance, although we have mentioned them in the Introduction.
The equation of state must have the
form of $P=k\rho$ from the requirement of self-similarity, 
where $k$ is a constant. The GRLP solution represents a naked-singularity formation
in the range $0<k\lesssim 0.0105$, where
the upper bound is imposed by the nakedness of the singularity. 
This solution is interesting because it provides the first 
example in which the pressure does not prevent 
the formation of a NS.
Moreover, the convergence to the GRLP solution of more general solutions
near the central region of stars have been strongly 
suggested numerically and supported by a mode analysis~\cite{HaradaMaeda} as a realization of the self-similarity hypothesis~\cite{Carr}. From the above, the GRLP solution
will be a strongest known counterexample against the cosmic censorship hypothesis. 

Because the GRLP solution is a numerical one, an
explicit expression of the exponent of local map could not obtained analytically,
although it is unlikely that $\alpha$ is equal to unity. 
To show rigorously that the power of emission is proportional to the inverse square of the remaining time to the CH and that the constant of proportion diverges in the limit to marginally NS, the dependence of exponent $\alpha$ on $k$ should be clarified numerically~\cite{preparation}.

\section{Summary and Discussion}
\label{sec:conclusion}
We have been concerned with a quantum mechanical particle creation
during the naked-singularity formation in spherically symmetric
self-similar collapse.
The power, energy, and redshift of emitted particles
are analytically calculated
on the assumption
that the curvature around the singularity causes particle creation and
the metric function $f$ is $C^{2-}$ around the CH.
As a result, in the generic case
in which the exponent of the local map $\alpha\neq 1$, the 
power has been found to diverge as $P\varpropto (t_{ch}-t)^{-2}$, where $(t_{ch}-t)$ is
the remaining time until a distant observer would receive
a first light ray from the NS.
It is worth pointing out that the weaker differentiability of the CH leads to
a different result, i.e., the power of emission has a
different time dependence, although we have not looked deeper
into such a possibility in this paper.
The square inverse proportion of the power to the remaining time
is due to the scale invariance of self-similar spacetimes.
The constant of proportion has been found to be arbitrarily large in the
limit to marginally NS.
Therefore, this explosive radiation is especially striking in the case that
the event horizon is very close to the CH because the emitted energy
can be arbitrarily large in spite of a cutoff expected from quantum gravity.
We go on from this to the conclusion that 
if the back reaction to a gravitational field is taken into account,
the semiclassical effect would cause the instability
of the CH and might recover the cosmic censor in this limiting case.
On the other hand, in the non-generic case in which 
$\alpha=1$ the power remains finite at the CH, so that
the semiclassical instability of the CH seems not to be efficient
for this special class of self-similar solutions.
The collapse of a massless scalar field described by the
Roberts solution indeed does correspond to this case. 
In addition, it has been found that the diverging redshift and blueshift
cause the divergence of the power to positive or negative infinity, depending on the manner
of the coupling of scalar fields to gravity.
The divergence will be a criterion for the stability/instability of a CH
in a gravitational collapse.
Having suggested that the particle creation could cause the instability of the CH,
we still have a long way to go before knowing the existence of
a classical/semiclassical cosmic censor in more generic spacetimes.

\acknowledgments
We are grateful to Kei-ichi Maeda for his continuous encouragement.
This paper owes much to the thoughtful and helpful comments of Hideki Maeda.
UM would like to thank L.H.~Ford for helpful comments. 
TH is also grateful to B.C.~Nolan and T.J.~Waters for helpful
discussions.
This work was partially supported by a Grant for The 21st Century COE Program (Holistic Research and Education Center for Physics 
Self-Organization Systems) at Waseda University.
TH was supported 
by a JSPS Postdoctoral Fellowship for Research Abroad.

\appendix

\section{First outgoing null ray from the singularity}
\label{sec:CH}
We shall prove 
the absence of an outgoing radial null geodesic which emanates from the origin 
$v=R=0$
before the null geodesic $x=x^{+}$.
Let $(v, R)$
be a point on a radial null geodesic $l$ in the
region $0<x=v/R<x^{+}$ which emanates from $v=R=0$.
Then, one can say that
\begin{eqnarray}
\left.\frac{dx}{dv}\right|_{l}=\frac{x}{v}(1-\frac{1}{f})>0.
\end{eqnarray}
By the uniqueness of the solution of the null geodesic equation, 
$l$ cannot cross $x=x^{+}$ at points other than $v=R=0$. 
Therefore above inequality says that
$x$ decreases as $v\to +0$ but is bounded from below by $0$.
Hence the limit 
\begin{eqnarray}
\bar{x}\equiv\left.\lim _{v\to +0}x(v)\right|_{l}
\end{eqnarray}
exists and satisfies $0\leq \bar{x}<x^{+}$. In the case of $0<\bar{x}<x^{+}$,
\begin{equation}
\bar{R}\equiv\left.\lim _{v\to +0}R(v)\right|_{l}\\
=\left.\lim _{v\to +0}\frac{v}{x(v)}\right|_{l}\\
=0
\end{equation}
holds, but
\begin{equation}
\bar{x}=\left.\lim _{v\to +0}\frac{v}{R(v)}\right|_{l}
=\left.\lim _{v\to +0}\frac{1}{R^{\prime}(v)}\right|_{l}
=\left.\lim _{v\to +0}xf(x)\right|_{l}=\bar{x}f(\bar{x}),
\end{equation}
where the l'Hopital's rule is used in the second equality. This contradicts
the assumption that $x^{+}$ is the smallest positive root of
equation $f(x)=1$. 
Next, we shall consider the case of $\bar{x}=0$.
The fact that the solution $R$ converges to a finite nonzero constant as $x\to 0$
from the condition (\ref{eq:condition_2})
contradicts the assumption that $l$ is a geodesic which 
emanates from $v=R=0$. Thus the case of $\bar{x}=0$ is also excluded.
Thus we see that $x=x^{+}$ is the first outgoing null ray which emanates from the singularity.

\section{Local Map and Redshift in Diagonal Coordinates}
\label{sec:Diagonal}
The line element of the class of spacetimes discussed in Sec. \ref{sec:spacetime} in a diagonal coordinate system is written as 
\begin{eqnarray}
ds^2=g_{tt}(z)dt^{2}+g_{rr}(z)dr^{2}+r^{2}S^{2}(z)d\Omega^{2},
\label{eq:metric_2}
\end{eqnarray}
where $z\equiv t/r$ and $S$ is a dimensionless metric function. The homothetic Killing vector field is of the form $\xi=t\partial _{t}+r\partial _{r}$. If one defines functions $w_{\pm}(z)$ as 
\begin{eqnarray}
w_{\pm}(z)\equiv\pm\frac{1}{z}\sqrt{-\frac{g_{rr}}{g_{tt}}},
\label{eq:wpm}
\end{eqnarray}
the roots of the algebraic equation $w_{\pm}(z)=1$ play important roles as do those of $f(x)=1$
in Sec.~\ref{sec:spacetime}.
The existence of a positive (negative) root of $w_{+}(z)=1$ ($w_{-}(z)=1$) and
the uniqueness of the root of $w_{-}(z)=1$ are assumed.
When the negative root and the smallest positive
root are denoted by $z^{-}$ and $z^{+}$ respectively,
the curves $z=z^{+}$ and $z=z^{-}$ can be shown to be the CH and 
the ingoing null ray that terminates at the NS.
In addition, it is assumed that $w^{\prime}_{\mp}(z^{\mp})\gtrless 0$.

The null geodesic equations are integrated to give
\begin{equation}
\frac{r}{r_{0}^{\pm}}=\exp\left[\int _{z_{0}^{\pm}}^{z}W_{\pm}(z^{\prime})dz^{\prime}\right]
=\exp\left[\int _{z_{0}^{\pm}}^{z}W^{\ast}_{\pm}(z^{\prime})dz^{\prime}\right]\left(\frac{z^{\pm}-z}{z^{\pm}-z_{0}^{\pm}}\right)^{1/\delta _{\pm}},
\label{eq:leading_2}
\end{equation}
where
\begin{eqnarray}
W_{\pm}(z)&\equiv&\frac{1}{z\left(w_{\pm}(z)-1\right)},\nonumber\\
\label{eq:Wpm}
\delta _{\pm}&\equiv&z^{\pm}w_{\pm}^{\prime}(z^{\pm}),\label{eq:gamma_pm}\\
W^{\ast}_{\pm}(z)&\equiv&W_{\pm}(z)-\frac{1}{\delta _{\pm}(z-z^{\pm})},\nonumber
\end{eqnarray}
and the signature $+$ $(-)$ corresponds to outgoing (ingoing) null geodesic.
The constants $z_{0}^{\pm}$ and $r_{0}^{\pm}$ are related as
$r_{0}^{+}=r(z=z_{0}^{+})$ for an outgoing ray, while
$r_{0}^{-}=r(z=z_{0}^{-})$ for an ingoing one,
where $z_{0}^{\pm}$ is set as $z_{0}^{\pm}<z^{\pm}$ and $z_{0}^{\pm}\neq 0$.
The constants $r^{\pm}_{0}$ are related to $t_{c}\equiv t(r=0)$ as
\begin{eqnarray}
r_{0}^{\pm}=-\frac{t_{c}}{|z_{0}^{\pm}|J_{\pm}},\;\;
J_{\pm}=\exp\left[\int _{z_{0}^{\pm}}^{-\infty}\frac{w_{\pm}(z^{\prime})dz^{\prime}}{z^{\prime}\left(w_{\pm}(z^{\prime})-1\right)}\right].
\label{eq:initial_2}
\end{eqnarray}
Combination of Eqs.~(\ref{eq:leading_2}) and (\ref{eq:initial_2}) yields
\begin{eqnarray}
r&=&D_{\pm}(r,z)\left(t^{\pm}(r)-t\right)^{1/\delta _{\pm}}t_{c},
\label{eq:radial_2}
\end{eqnarray}
where
\begin{eqnarray}
t^{\pm}(r)&\equiv&z^{\pm}r,\nonumber\\ 
D_{\pm}(r,z)&\equiv&-|z_{0}^{\pm}|^{-1}J_{\pm}^{-1}\left[(z^{\pm}-z_{0}^{\pm})r\right]^{-1/\delta _{\pm}}\exp\left[\int _{z_{0}^{\pm}}^{z}W_{\pm}^{\ast}(z^{\prime})dz^{\prime}\right].\nonumber
\label{eq:C_2}
\end{eqnarray}

Now, let us consider a pair of ingoing and outgoing null rays such
that the latter is the reflection of the former at the regular center $(r=0, t<0)$. 
An observer who rests at $r=\mathfrak{r}$ will encounter the null ray twice, so that
we denote the time of first encounter by $t_{1}$ and that of the second by $t_{2}$. By Eq.~(\ref{eq:radial_2}), the ingoing and outgoing null rays are matched at the center to give the local map as
\begin{eqnarray}
t^{-}(\mathfrak{r})-t_{1}=\left[\frac{D_{+}(\mathfrak{r},z_{2})}{D_{-}(\mathfrak{r},z_{1})}\right]^{\delta _{-}}
\left(t^{+}(\mathfrak{r})-t_{2}\right)^{\alpha _{2}},
\label{eq:mapping_2}
\end{eqnarray} 
where 
\begin{eqnarray}
z_{i}&\equiv&t_{i}/\mathfrak{r},\,\,(i=1,2),\nonumber \\
\alpha _{2}&\equiv&\delta _{-}/\delta _{+}.
\label{eq:alpha_diagonal}
\end{eqnarray}

The redshift of a radial null ray is obtained in similar way.
The $t$-component of equation $k^{\mu}\nabla _{\mu}k^{\nu}=0$ is integrated to give
\begin{equation}
\frac{k^{t}(z)}{k^{t,\pm}_{0}}=\exp\left[\int _{\tilde{z}_{0}^{\pm}}^{z}\tilde{W}_{\pm}(z^{\prime})dz^{\prime}\right]
=\exp\left[\int _{\tilde{z}_{0}^{\pm}}^{z}\tilde{W}_{\pm}^{\ast}(z^{\prime})dz^{\prime}\right]\left(\frac{z^{\pm}-z}{z^{\pm}-\tilde{z}_{0}^{\pm}}\right)^{-(1+\delta _{\pm})/\delta _{\pm}},\label{eq:leading_k2}
\end{equation}
where 
\begin{eqnarray}
\tilde{W}_{\pm}(z)&\equiv&-\frac{1}{2\left(1-w_{\pm}^{-1}\right)}\left\{\left(1-2w^{-1}_{\pm}\right)\frac{1}{g_{tt}}\frac{dg_{tt}}{dz}+\frac{1}{g_{rr}}\frac{dg_{rr}}{dz}\right\},\nonumber\\
\label{eq:tilde_Wpm}
\tilde{W}_{\pm}^{\ast}(z)&\equiv& \tilde{W}_{\pm}(z)+\frac{1+\delta _{\pm}}{\delta _{\pm}}\frac{1}{z-z^{\pm}}.\nonumber
\end{eqnarray}
The constants $\tilde{z}_{0}^{\pm}$ and $k^{t,\pm}_{0}$ are related as $k^{t,+}_{0}=k^{t}(z^{+}_{0})$ for an outgoing ray, while $k^{t,-}_{0}=k^{t}(z^{-}_{0})$ for an ingoing one.
The constants $k^{t,\pm}_{0}$ are related to $k^{t}_{c}\equiv k^{t}(r=0)$ as 
\begin{eqnarray}
k^{t,\pm}_{0}=\frac{k^{t}_{c}}{\tilde{J}_{\pm}},\;\;
\tilde{J}_{\pm}\equiv\exp\left[\int _{\tilde{z}_{0}^{\pm}}^{-\infty}\tilde{W}_{\pm}(z^{\prime})dz^{\prime}\right].
\label{eq:initial_k2}
\end{eqnarray}
Combination of Eqs.~(\ref{eq:leading_k2}) and (\ref{eq:initial_k2}) yields
\begin{eqnarray}
k^{t}(z)=\tilde{D}_{\pm}(r, z)\left(t^{\pm}(r)-t\right)^{-(1+\delta _{\pm})/\delta _{\pm}}k^{t}_{c},
\label{eq:k_2}
\end{eqnarray}
where
\begin{eqnarray} 
\tilde{D}_{\pm}(r,z)\equiv\tilde{J}_{\pm}^{-1}\left[(z^{\pm}-\tilde{z}_{0}^{\pm})r\right]^{(1+\delta _{\pm})/\delta _{\pm}}\exp\left[\int _{\tilde{z}_{0}^{\pm}}^{z}\tilde{W}^{\ast}_{\pm}(z^{\prime})dz^{\prime}\right].\nonumber
\end{eqnarray}
Consider again
the observer who rests at $r=\mathfrak{r}$ and the pair of ingoing and outgoing null rays.
The outgoing and ingoing null rays are matched at the center by 
Eqs.~(\ref{eq:mapping_2}) and (\ref{eq:k_2}) to give 
\begin{eqnarray}
\frac{\hat{\omega}_{2}}{\hat{\omega}_{1}}=\sqrt{\left|\frac{g_{tt}(z^{+})}{g_{tt}(z^{-})}\right|}\frac{\tilde{D}_{+}(\mathfrak{r}, z_{2})}{\tilde{D}_{-}(\mathfrak{r}, z_{1})}\left[\frac{D_{+}(\mathfrak{r}, z_{2})}{D_{-}(\mathfrak{r}, z_{1})}\right]^{1+\delta _{-}}\left(t^{+}(\mathfrak{r})-t_{2}\right)^{\alpha _{2}-1},
\label{eq:total_redshift_3}
\end{eqnarray}  
where $\hat{\omega}_{1}\equiv\lim_{z_{1}\to z^{-}}\sqrt{|g_{tt}(z_{1})|}\,k^{t}(z_{1})$ and $\hat{\omega}_{2}\equiv\lim_{z_{2}\to z^{+}}\sqrt{|g_{tt}(z_{2})|}\,k^{t}(z_{2})$ are the observed frequencies.

There exists a plausible relation between the local map and redshift.
From Eqs.~(\ref{eq:mapping_2}) and (\ref{eq:total_redshift_3}), one obtains
\begin{eqnarray*}
\frac{d\tau_{2}}{d\tau_{1}}=\frac{\hat{\omega}_{1}}{\hat{\omega}_{2}},
\end{eqnarray*}
where $d\tau _{i}\equiv\sqrt{|g_{tt}|}\,dt_{i}$ ($i=1,2$) is the proper time of the observers.



\begin{thebibliography}{99}
\bibitem{footnote_2}
One can prove $f^{\prime}(x^{+})\leq 0$ with the dominant energy condition on collapsing matter and the equality is excluded by a physically reasonable requirement on the collapsing matter at the CH~\cite{Nolan}.
\bibitem{footnote_3}
The time $t_{\mathrm{QG}}$ can be regarded as the Planck time $t_{\mathrm{pl}}$;
at least this is the case for self-similar dust model.
\bibitem{footnote_4}
When the CH and event horizon exactly coincide, radiation reduces
to the Hawking one~\cite{Ford,Hiscock}.
This fact cannot be derived with the method making use of the local map since
in this case, $\gamma=0$ in Eq.~(\ref{eq:leading_1}).
It is not surprising since an event horizon, which plays a central role
in the Hawking radiation, is not a local object but a global one.
\end{thebibliography}
\end{document}